\documentclass[preprint,showpacs,preprintnumbers,amsmath,amssymb]{revtex4}
\usepackage{graphicx}% Include figure files
\usepackage{dcolumn}% Align table columns on decimal point

\begin{document}

\title{On the nature of two superconducting transitions in the specific heat of PrOs$_4$Sb$_{12}$: Effects of crystal grinding}

\author{M.\ E.\ McBriarty}
\affiliation{ Department of Physics, University of Florida
P.O.\ Box 118440, Gainesville, Florida 32611--8440, USA}

\author{P.\ Kumar}
\affiliation{ Department of Physics, University of Florida
P.O.\ Box 118440, Gainesville, Florida 32611--8440, USA}

\author{G.\ R.\ Stewart}
\affiliation{ Department of Physics, University of Florida
P.O.\ Box 118440, Gainesville, Florida 32611--8440, USA}

\author{B.\ Andraka}
\email{andraka@phys.ufl.edu}
\affiliation{ Department of Physics, University of Florida
P.O.\ Box 118440, Gainesville, Florida 32611--8440, USA}

\date{\today}

\begin{abstract}
Specific heat, dc- and ac-magnetic susceptibility are reported for a large single crystal of PrOs$_4$Sb$_{12}$ and, after grinding, its powder. The room temperature effective paramagnetic moment of the crystal was consistent with the Pr$^{3+}$ ionic configuration and full occupancy of the Pr-sublattice. The crystal showed two distinct anomalies in the specific heat and an overall discontinuity in $C/T$ of more than 1000 mJ/K$^2$mol. The upper transition (at $T_{c1}$) was rounded, in an agreement with previous reports. The anomaly at $T_{c2}$ was very sharp, consistent with a good quality of the crystal. We observed a shoulder in $\chi$' and two peaks in $\chi$'' below $T_{c1}$. However, there were no signatures in $\chi$' of the lower temperature transition. PrOs$_4$Sb$_{12}$ is extremely sensitive to grinding, which suppresses the upper superconducting transition in both the specific heat and magnetic susceptibility. $\Delta C/T_{c}$ was reduced to 140 mJ/K$^2$ mol in the powdered sample. Existing data on ground, polished, and sliced crystals suggests the existence of a length scale of order 100 $\mu$, characterizing the higher temperature superconducting phase. 
\end{abstract}


\maketitle

\section{Introduction}

PrOs$_4$Sb$_12$ is the first discovered Pr-based heavy fermion superconductor\cite{Bauer} that attracts widespread attention on account of several anomalous properties. These anomalies include a multiplicity of ground states and transitions and unusual signatures of these transitions. Most of the reported samples exhibited two superconducting transitions\cite{Maple,Vollmer,Cichorek,Rotundu,Measson,Measson2,Grube} accompanied by huge discontinuities in the specific heat and a holding drop in the magnetic susceptibility as a function of temperature below $T_c$. The last feature would lead one to suspect that the appearance of coherence in this system is inhomogeneous, i.e. the flux expulsion takes place in a non-uniform manner. Measson et al.\cite{Measson} have reported a decrease in the ratio of the specific heat discontinuities at the upper and lower transitions in a sample that was reduced to a size of 120 $\mu$ by polishing, arguing further in support of inhomogeneous ground states and sensitivity to defects.  It has been suggested that the upper transition may be associated with Pr-deficient regions near a surface.
 
To further investigate these anomalies, we report here results on a 
sample that has been first measured for its properties, such as specific heat, ac- and dc-magnetic susceptibility, and then, for 
comparison, ground to dimensions of order 20-100 $\mu$ and measured again.  We find unusually strong effects of the grinding on superconducting anomalies as well as the normal state specific heat.

\section{Single crystal}

The investigated crystal ($\approx 20$ mg) had almost perfect cubic shape with all six faces smooth and regular. It was obtained by slow, 1 deg/h, cooling of PrOs$_4$Sb$_{20}$ (the "self-flux" method).  Before presenting the superconducting characteristics of the crystal we discuss its magnetic susceptibility measured in a Quantum Design squid magnetometer between 1.8 and 350 K.  The measurement was performed using a configuration\cite{straw} to minimize holder contribution to the magnetization, which in the case of small single crystals can be larger than the magnetization of the sample itself. The inset to Fig. 1 shows the susceptibility ($\chi$) in a field of 0.5 T parallel to the (001) crystallographic direction. The susceptibility has a low temperature maximum of 112 memu/mol at 3.5 K. The paramagnetic effective moment, obtained from the straight line fit of  $1/\chi$ versus $T$ (temperature) between 200 and 350 K, is 3.63 $\mu_B$/Pr (Fig. 1).  This fit results in a very small, approximately 1 K, positive Curie-Weiss temperature. A similar fit performed for the data between 150 and 300 K yields an effective moment of 3.65 $\mu_B$/Pr and a negative Curie-Weiss temperature of 3 K. The small absolute value of the derived Curie-Weiss temperatures indicates that Pr-moments are essentially non-interacting with each other at these high temperatures. These high temperature effective moments are only slightly larger than the theoretical value for free trivalent Pr of 3.58 $\mu_B$.  Considering this (weak) variation of the measured moment with fitting range, we might expect that at temperatures much higher than the overall crystal field splitting (210 K) the measured value would be even closer to the theoretical one. At this point we stress that susceptibility measurements on all our single crystals with mass of at least 5 mg resulted in almost identical values of the high temperature effective moment. This is in disagreement with most of the published magnetic susceptibility results\cite{Bauer,Maple,Grube} by other research groups implying much smaller $\mu_{eff}$ or a large concentration of Pr-vacancies. The susceptibility results suggest that the Pr occupancy of our crystal is close to 1.

\begin{figure}
\includegraphics[width=3.0in]{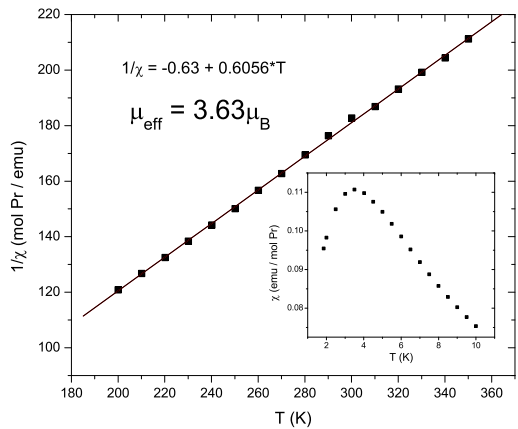}
\caption{\label{susceptibility}%
Inverse of magnetic susceptibility versus temperature for the single crystal of PrOs$_4$Sb$_{12}$. The inset shows low temperature magnetic susceptibility measured in 5 kOe along the (001) direction.}
\end{figure}

The specific heat data between 1.45 and 2 K are shown in Fig. 2. Our crystal exhibits a clearer separation of the two specific heat anomalies than the majority of previously published results (upper panel). The onset of the first superconducting transition is at 1.84 K ($T_{c1}$) both in the specific heat and in the $C/T$ (specific heat divided by temperature).  $C$ has a weak maximum near 1.79 K. The onset of the second transition is at about 1.73 K ($T_{c2}$) and $C$ ($C/T$) has a maximum at 1.7 K. Thus, as opposed to the upper transition, the anomaly corresponding to $T_{c2}$ is very sharp suggesting good quality of the crystal, consistent with the excellent separation of the transitions. During our measurement the specific heat was averaged over at least 2\% of $T$ (e.g., over $\approx$ 40 mK at 1.8 K). Thus, if not for this averaging procedure (necessary to avoid large scattering) we would expect this transition to be even sharper. $C/T$ reaches a maximum near 1.7 K. The overall $\Delta C/T$ (without the usual conservation of entropy construction, which would increase this value) is approximately 1000 mJ/K$^2$mol, among the largest ever reported. The height of the first anomaly in $C/T$ is roughly 1/2 of the anomaly at $T_{c2}$.  Nevertheless, both transitions are clearly bulk transitions. If they corresponded to different regions of the crystal (inhomogeneous scenario) we would expect a somewhat larger region of the material undergoing the transition at $T_{c2}$, although both regions are comparable. 

\begin{figure}
\includegraphics[width=3.0in]{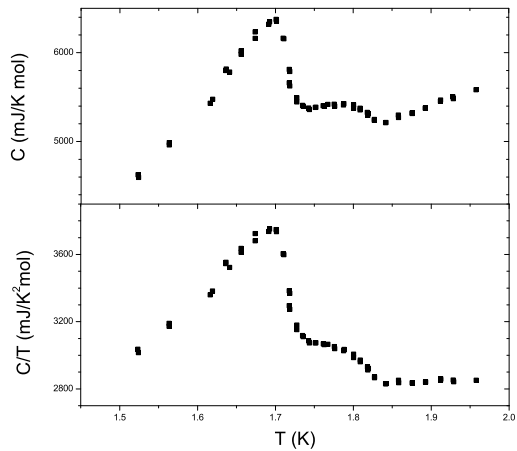}
\caption{\label{Cp:crystal}%
Specific heat (upper panel) and specific heat divided by temperature (lower panel) versus temperature for the single crystal of PrOs$_4$Sb$_{12}$.} \end{figure}

The ac-susceptibility measurement performed on the same crystal is shown in Fig. 3. The upper panel is the real part of the ac-susceptibility ($\chi$'), the lower panel is the imaginary part ($\chi$'').  The temperature variation of the real part of the susceptibility in a typical superconductor is step-like due to a transition from a full penetration of the ac-field at higher temperatures to a perfect screening of this field below $T_c$.  The imaginary part is due to dissipation associated with flux motion. The onset of the superconducting transition is approximately at 1.85 K according to $\chi$'. This corresponds well to the onset of the bulk transition in either $C$ or $C/T$ shown in Fig. 2. $\chi$' versus $T$ exhibits a shoulder near 1.8 K. Such a shoulder could be due to a superposition of two steps related to two superconducting transitions. However, this possible lower temperature step takes place near 1.8 K, thus at a temperature higher than the onset of the lower susperonducting transition in $C$. A similar shoulder at 1.8 K (and an additional step at lower temperatures) can be seen for one of the crystals (n1b) reported in reference \cite{Measson2}. According to our $\chi$' data, the transition is essentially complete at 1.73 K, which is the onset of the lower superconducting transition in $C$ ($C/T$). Therefore, we have examined the dissipative part of the ac-susceptibility. Interestingly, our crystal showed two peaks in $\chi$'', at 1.84 and 1.77 K. Again, there is no structure in $\chi$'' below 1.73 K, corresponding to the lower superconducting transition in the specific heat. 

\begin{figure}
\includegraphics[width=3.0in]{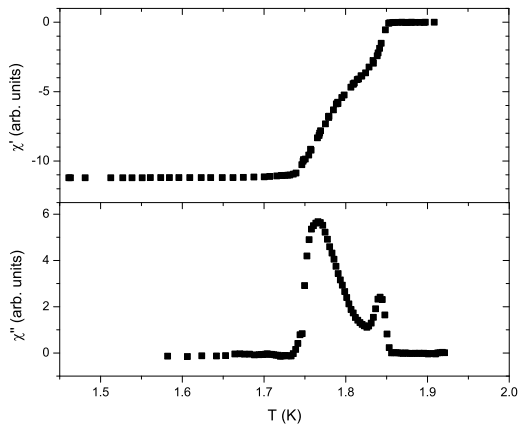}
\caption{\label{chi:ac}%
Real (upper panel) and imaginary (lower panel) parts of the ac-susceptibility for the single crystal of PrOs$_4$Sb$_{12}$.} \end{figure}

The two peak-effect in $\chi$'' has been previously observed in some (high temperature) granular superconductors investigated in superimposed ac- and dc- fields of approximately similar magnitude\cite{Gjolmesli}. The interpretation of the two-peak effect is that one of these peaks is due to the intragrain loss, the other is due to intergrain loss. The temperatures of the two peaks in granular superconductors is a strong function of ac- and dc-fields. Therefore, we have investigated another single crystal of PrOs$_4$Sb$_{12}$ (also approximately 20 mg), showing a two-peak effect in the ac-susceptibility, as a function of frequency and amplitude of ac-field (in Earth's dc-magnetic field). Detailed results of this study will be published elsewhere. We have found no frequency dependence of the temperatures of the two peaks. Contrary to granular superconductors, there is no variation of these two temperatures on the amplitude of the ac-field either, which was varied by a factor of 10. For sufficiently small values of the driving ac-field, the two peaks are very sharp defining two characteristic temperatures of the material. These temperatures correspond to the steepest drops of $\chi$' versus $T$. Interestingly, this upper temperature peak was at 1.84 K for both crystals. 

The separation of these two peaks in the ac-susceptibility is $70 \pm 10$ mK, smaller than the difference of $T_{c1}$ and $T_{c2}$ defined by the onset of the anomalies in the specific heat. However, the onset of the anomaly corresponding to $T_{c2}$ might be hindered by the higher temperature superconducting transition. It is interesting that we do not observe a step in $\chi$' associated with the very sharp peak at $T_{c2}$. Recall that the crystal investigated by Cichorek and collaborators\cite{Cichorek} also did not show any signature of the lower superconducting transition in the ac-susceptibility. Obviously, this can be explained by the shielding effect of the regions becoming superconducting at $T_{c1}$.  The results of Measson et al.\cite{Measson} suggest that the material near the surface of formed single crystals might have a larger $T_{c}$ than the interior of the crystal.

\section{Powder}

In order to break the connectivity of possibly higher $T_{c}$ material near the surface of the crystal, we powdered our crystal. The powdering was done in an agate mortar to avoid any magnetic contamination.  No measurement of the size of the grains was performed for this crystal. We have checked the size of grains of a smaller crystal, powdered in the same manner. Almost all the powder was in the 20 - 100 $\mu$m range. Thus, these grains are larger  by at least a factor of 1000 than the superconducting coherence length (170 \AA)\cite{Bauer}. Most of this powder was subsequently pressed into a 1/8-in. pellet. A very small amount of the (unpressed) powder was mixed with GE varnish and attached to a copper screw for the ac-susceptibility measurement. 

We have observed small effects of the powdering on the dc-magnetic susceptibility up to 10 K. The low temperature maximum shifts by about 0.5 K (not shown) to a lower temperature and becomes more rounded. We have not attempted to extract the high temperature effective moment, since a crystal holder, with a relatively large magnetization at room temperature, was used. 

\begin{figure}
\includegraphics[width=3.0in]{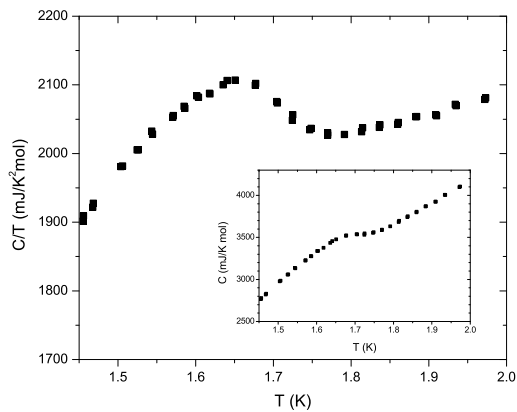}
\caption{\label{Cp:powder}%
Specific heat divided by temperature (lower panel) versus temperature for the powder of PrOs$_4$Sb$_{12}$. The inset shows the specific heat for the powder of PrOs$_4$Sb$_{12}$.} \end{figure}

There is a profound effect of grinding on the specific heat (Fig. 4).  There is only a single specific heat anomaly in the powdered material. The onset of the broad transition is at approximately 1.76 K. By performing a conservation of entropy construction we arrive at $T_{c}$ of approximately 1.71 K, which is close to $T_{c2}$ for the single crystal. $\Delta C/T$ obtained by the same procedure is 140 mJ/K$^2$mol (as compared to 1000 mJ/K$^2$mol for the crystal).  We do not observe any signature corresponding to $T_{c1}$. Since the specific heat is a bulk property, the specific heat technique might not be sensitive enough to detect a small amount of material becoming superconducting at $T_{c1}$. 

\begin{figure}
\includegraphics[width=3.0in]{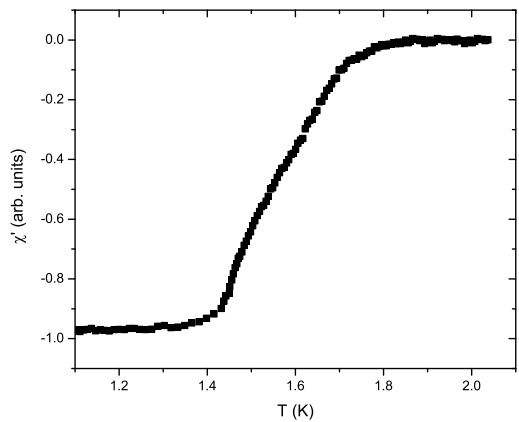}
\caption{\label{chi:acpow}%
ac-susceptibility versus temperature for the powder of PrOs$_4$Sb$_{12}$.} \end{figure}

The results of the real part of the ac-susceptibility measurement are presented in Fig. 5. There seems to be only a very weak signature of the onset of superconductivity at 1.85 K. However, the decrease in $\chi$' between 1.85 and 1.72 K is less than 10\% of the drop in $\chi$' taking place below 1.72 K. The $\chi$' transition is very wide, in comparison with that of the single crystal, and is not complete until approximately 1.4 K. Considering the fact that the ac-susceptibility technique should be more sensitive in detecting the higher than the lower $T_{c}$, we can safely assume that the amount of material superconducting above 1.72 K is small, much smaller than 10\% of the total. Thus, grinding suppresses preferentially the upper $T_{c}$ phase. Within the resolution of our measurement we have not seen any structures in $\chi$'' in our powder.  

\section{Discussion}

The most interesting observation for the single crystal is an apparent disconnection between various characteristics of the specific heat and ac-susceptibility. The specific heat exhibits a very sharp peak at $T_{c2}$, suggesting a high quality of the crystal, but a rounded anomaly at $T_{c1}$. This higher temperature feature can be understood in terms of inhomogeneous superconductivity\cite{Measson}, where there is a distribution of $T_{c}$'s due to a varying Pr-stoichiometry. However, the $\chi$'' has a very sharp peak in low excitation fields at a specific temperature of 1.84 K (this temperature does not seem to be crystal-dependent either) suggesting a lack of a significant distribution of $T_{c1}$. $\chi$' has a shoulder at 1.8 K, i.e. at too high temperature to be associated with the lower temperature superconducting transition. The temperature variation of $\chi$' is very similar to the change of the penetration depth near $T_{c}$ measured by Chia et al.\cite{Chia} Obviously, the two properties are closely related. The penetration depth also has a shoulder at 1.8 K, which is at a significantly larger temperature than $T_{c2}$. The origin of this shoulder in both properties, confirmed by us on other large crystals by ac-susceptibility and seen in other published results\cite{Measson2}, is unknown. 

Both the specific heat and magnetic susceptibility data imply that the superconductivity of PrOs$_4$Sb$_{12}$ is very sensitive to crystal grinding. Grinding breaks the connectivity of possibly different crystallographic phases of the crystal, but also introduces some amount of strain and defects. It can also lead to size effects. As it was already mentioned, the size of grains is still much larger than the coherence length in a single crystal\cite{Bauer}. However, one should keep in mind that the coherence length was calculated from $dH_{c2}/dT$ at $T_c$, while $H_{c2}$ versus $T$ is strongly non-linear near $T_{c}$ in PrOs$_4$Sb$_{12}$. Very recently, Seyfarth\cite{Seyfarth} and collaborators observed a disappearance of the upper superconducting anomaly in the specific heat when a crystal was sliced to a platelet 50 $\mu$m thick. This thickness is comparable to the size of grains of our powdered sample. The main difference between our results for powdered sample and those of Seyfarth at al. is that the specific heat anomaly at $T_{c2}$ in the sliced crystal is as sharp as in the original crystal.   

\begin{figure}
\includegraphics[width=3.0in]{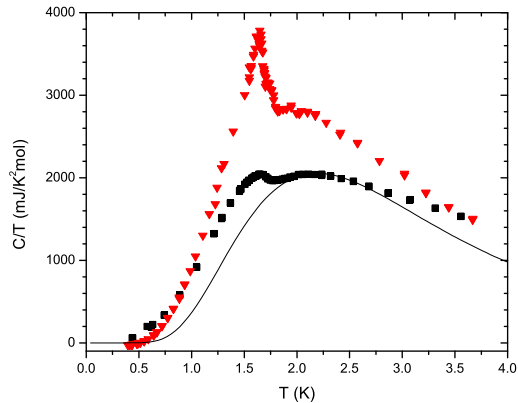}
\caption{\label{Cp:powder2}%
(Color on-line) $f$-electron specific heat of a single crystal (triangles) and powdered (squares) PrOs$_4$Sb$_{12}$ between 0.4 and 4 K. Non-$f$-electron contributions were accounted for using the normal-state specific heat data of LaOs$_4$Sb$_{12}$. The solid line is a Schottky specific heat, corresponding to the system with a singlet ground state separated from the excited doublet by 7.8 K, multiplied by 0.59.}\end{figure}

We believe that powdering leads to a distribution of crystalline electric field (CEF) parameters as suggested by the broadening of the low temperature maximum in the magnetic susceptibility. This broadening becomes even more apparent by inspecting the specific heat above $T_{c}$. $C/T$ at 1.95 K (near Schottky maximum) is reduced from over 2800 mJ/K$^2$mol for the crystal to 2070 mJ/K$^2$mol for the powder (Fig. 2 and 4). Since the crystal, which was powdered, was investigated by specific heat up to 2 K only, in Fig. 6 we compare the $f$-electron $C/T$ (obtained by subtracting normal electrons and phonons contributions using the LaOs$_4$Sb$_{12}$ data\cite{Rotundu}), between 0.4 and 4 K, of the powder and another crystal. Both crystals showed very good overlap of $C/T$ below 2 K. The solid line represents a Schottky $C/T$ for excitations between a singlet (ground) and triplet (excited) states, multiplied by a factor of 0.59, in order to match the $C/T$ values at 2.2 K for the powder. Clearly, the experimental Schottky-like anomaly in the powder is much wider than that expected for the accepted CEF model.\cite{Aoki,Kohgi,Andraka,Goremychkin,Ingersent} This experimental maximum is also much wider than that in Ru-doped\cite{Frederick2} or strongly La-doped crystals.\cite{unpublished} Because of a weak binding of rare-earths in the skutterudites, we expect a large number of dislocation of Pr-ions from their equilibrium positions upon grinding. It is interesting to note that the experimental Schottky anomaly in single crystals is narrower than the theoretical maximum describing CEF excitations, suggesting importance of additional electronic degrees of freedom in the system. 

Our results on a single crystal suggest a possibility that unusual structures in the ac-suceptibility are associated with the broad upper transition and are not related to the transition at $T_{c2}$. The extreme sensitivity of the transition at $T_{c1}$ to grinding does not seem to be consistent with the hypothesis of structural inhomogeneities responsible for the transition at $T_{c1}$. Grinding, after all, introduces all kind of structural inhomogeneities and should therefore lead to a relative enhancement of the anomaly at $T_{c1}$ with respect to that at $T_{c2}$. The rounded anomaly in the specific heat and small initial rate of flux expulsion (ac-susceptibility) allow to speculate that the transition of $T_{c1}$ is of a higher (than two) order phase transition. This upper temperature transition is extremely sensitive to any modifications of the sample, such as powdering or slicing to a small thickness. Our results, together with those of Measson and collaborators\cite{Measson,Measson2}, suggest a possibility of an existence of an additional length scale of order 100 $\mu$, characterizing the high temperature superconducting state. 

Finally, our work points to significant differences between single crystals and powder of PrOs$_4$Sb$_{12}$. Since experimental techniques that use powdered material, such as NMR or neutron diffraction, have contributed to our current understanding of this system, those results should be carefully re-evaluated. 

\begin{acknowledgments}
This work has been supported by U.S. Department of Energy (DOE) Grant
Numbers DE-FG02-99ER45748 (M.R.M. and B.A.) and DE-FG02-86ER45268 (G.R.S.). 

\end{acknowledgments}


\begin{thebibliography}{19}
\bibitem{straw}
The crystal was placed between two concentric straws of the same length, which was much larger than the length of an individual scan.
\end{thebibliography}
\end{document}